\documentclass[conference]{IEEEtran}
\IEEEoverridecommandlockouts
\usepackage{cite}
\usepackage{amsmath,amssymb,amsfonts}
\usepackage{algorithmic}
\usepackage{graphicx}
\usepackage{textcomp}
\usepackage{xcolor}
\usepackage{hyperref}
\usepackage{array}
\def\BibTeX{{\rm B\kern-.05em{\sc i\kern-.025em b}\kern-.08em
    T\kern-.1667em\lower0.7ex\hbox{E}\kern-.125emX}}
\begin{document}

\title{Bug Destiny Prediction in Large Open-Source Software Repositories through Sentiment Analysis and BERT Topic Modeling}

\author{
\IEEEauthorblockN{Sophie C. Pope\textsuperscript{*}}
\IEEEauthorblockA{\textit{Department of Computer Science} \\
\textit{University of Colorado Colorado Springs}\\
\textit{United States} \\
\textit{and University of Wisconsin-La Crosse, WI, United States}\\
Poppysc@icloud.com} ~\\
\and
\IEEEauthorblockN{Andrew Barovic and Armin Moin\textsuperscript{*}}
\IEEEauthorblockA{\textit{Department of Computer Science} \\
\textit{University of Colorado Colorado Springs}\\
\textit{United States} \\
{\{abarovi2, amoin\}}@uccs.edu \\
\textsuperscript{*}Corresponding author
}
}
\maketitle

\begin{abstract}
This study explores a novel approach to predicting key bug-related outcomes, including the time to resolution, time to fix, and ultimate status of a bug, using data from the Bugzilla Eclipse Project. Specifically, we leverage features available before a bug is resolved to enhance predictive accuracy. Our methodology incorporates sentiment analysis to derive both an emotionality score and a sentiment classification (positive or negative). Additionally, we integrate the bug's priority level and its topic—extracted using a BERTopic model—as features for a Convolutional Neural Network (CNN) and a Multilayer Perceptron (MLP). Our findings indicate that the combination of BERTopic and sentiment analysis can improve certain model performance metrics. Furthermore, we observe that balancing model inputs enhances practical applicability, albeit at the cost of a significant reduction in accuracy in most cases. To address our primary objectives—predicting time-to-resolution, time-to-fix, and bug destiny—we employ both binary classification and exact time value predictions, allowing for a comparative evaluation of their predictive effectiveness. Results demonstrate that sentiment analysis serves as a valuable predictor of a bug’s eventual outcome, particularly in determining whether it will be fixed. However, its utility is less pronounced when classifying bugs into more complex or unconventional outcome categories.
\end{abstract}

\section{Introduction} \label{introduction}
Large open-source projects offer an issue tracking system or open bug repository where developers and users can report the software defects they find or any new feature requests they may have. These reports are called \textit{bug reports}. We focused on the natural language data available in descriptions of bug reports when they are reported by users and the priority of a bug report. The priority of a bug report is the assigned importance of a bug from the developers. For example, in the eclipse bug project, there are five levels: P1-P5. P1 indicated that the bug is very severe, and P5 meant that the bug is not severe at all. An example of a P1 bug is if the software crashed, and users were completely unable to use it.

Moreover, developers often use revision control systems (version control systems or source code repositories), such as Git. When they want to commit code to these repositories, they often write a commit log or commit message. These logs or messages also included valuable natural language data. In some cases, they also include the unique identifier (ID) of the bug report to which this specific change in the code base is related. For instance, if they managed to resolve an issue, they typically include the bug ID in their commit log.

We applied recent advances in Natural Language Processing (NLP) to analyze the \textit{sentiments} of the above-mentioned natural language text (i.e., bug reports and commit logs) and explore possible correlations between the detected sentiments and the bug reports, and how much time it took to fix the bugs, i.e., \textit{time-to-fix}. We also looked at \textit{time-to-resolution}. The difference between these is that \textit{time-to-resolution} referred to the time until a conclusion of a bug is reached, which included labels such as \textit{won't-fix}, \textit{invalid}, \textit{works-for-me}, \textit{duplicate}, \textit{Not\_Eclipse}, and \textit{nduplicate}. In contrast, \textit{time-to-fix} only applies to bug reports whose resolution is stated as fixed. We proposed that predicting \textit{time-to-resolution} of a bug is more useful in practice because we cannot know absolutely whether a reported bug will be fixed. Therefore, to make our analysis more practical, we focused on \textit{time-to-resolution} and did not exclude the bugs that were not fixed. We discuss this further in the \ref{related-work} section.

Also explained in Section \ref{related-work}, prior works in the literature have applied Natural Language Processing (NLP) to software repositories. This often falls under the broad topic of Mining Software Repositories (MSR). For instance, AI-based approaches have been used to address the bug triage problem. Examples include using Machine Learning (ML) \cite{Anvik+2006} (including graph learning \cite{Jeong+2009}) and Information Retrieval (IR) \cite{Sun+2011} to (semi-)automate bug triage. \textit{Bug triage} is the process of finding invalid and duplicate bug reports and assigning the valid and unique ones to the developers working on the project so they can fix and resolve them. In large open-source projects a group of developers, called \textit{triagers}, are typically responsible for the bug triage task. 

By predicting the time-to-resolution for bug reports, we can enable triagers to assign bugs that are prone to be time-consuming to developers with a higher rate of availability or who are known to be committed to the project for a longer period. This can reduce the likelihood of \textit{bug tossing} (i.e., reassignment of bug reports to other developers), which is costly for the project.

In this paper, we proposed a novel approach to automated bug destiny prediction using sentiment analysis and BERTopic from natural language text available in software bug and code repositories. We aimed to predict how long it would take developers to resolve a bug. Sentiment analysis has already been applied to bug repositories. For instance, Umer et al. \cite{Umer+2018} analyzed \textit{emotions} in bug reports to predict the priority level of the report. We also analyze the emotional sentiments in the descriptions in bug reports. However, we do this to predict the bug report's time-to-resolution (we call them the bug \textit{destiny}).

The contribution of this paper is twofold: First, it proposed and implemented a novel approach to automated bug report destiny prediction. This is validated by our experimental results using available data from open reference datasets deployed in the related work in the literature. Second, it provides an open-source prototype that enables other open-source projects using similar issue-tracking and revision control systems to use the proposed approach, thus benefiting from effective and efficient automated bug report destiny prediction.

This paper is structured as follows: Section \ref{background} provides some background information about open bug repositories and NLP. Further, we review the literature in Section \ref{related-work}. In Section \ref{proposed-approach}, we propose our novel approach and report on our experimental results in Section \ref{experimental-results}. Moreover, Section \ref{discussion} discusses the results and points out potential threats to validity. Finally, we conclude and suggest future work in Section \ref{conclusion-future-work}.

\section{Background}\label{background}
\subsection{Open Bug Repositories}
These repositories are open to users, and they contain reports about software issues or enhancements and track the progress \cite{anvik+2005}. These bugs can include wanted enhancements or issues with the software. We use these open bug repositories to analyze bug reports' destinies, including eventual status and time to resolution. As seen in Figure 1, the destiny of bugs will be entered where the status is and will change from new to fixed, \textit{won't-fix}, \textit{invalid}, or \textit{works-for-me}. Predicting the eventual status of a bug to help semi-automate the triage of bugs will save developers time and money because they will not be tasked to fix a bug that is predicted not to be fixable. Examples of bugs whose outcomes are not fixed are duplicate bugs or user errors. 
\begin{figure}[h]
    \centering
    \includegraphics[width=1\linewidth]{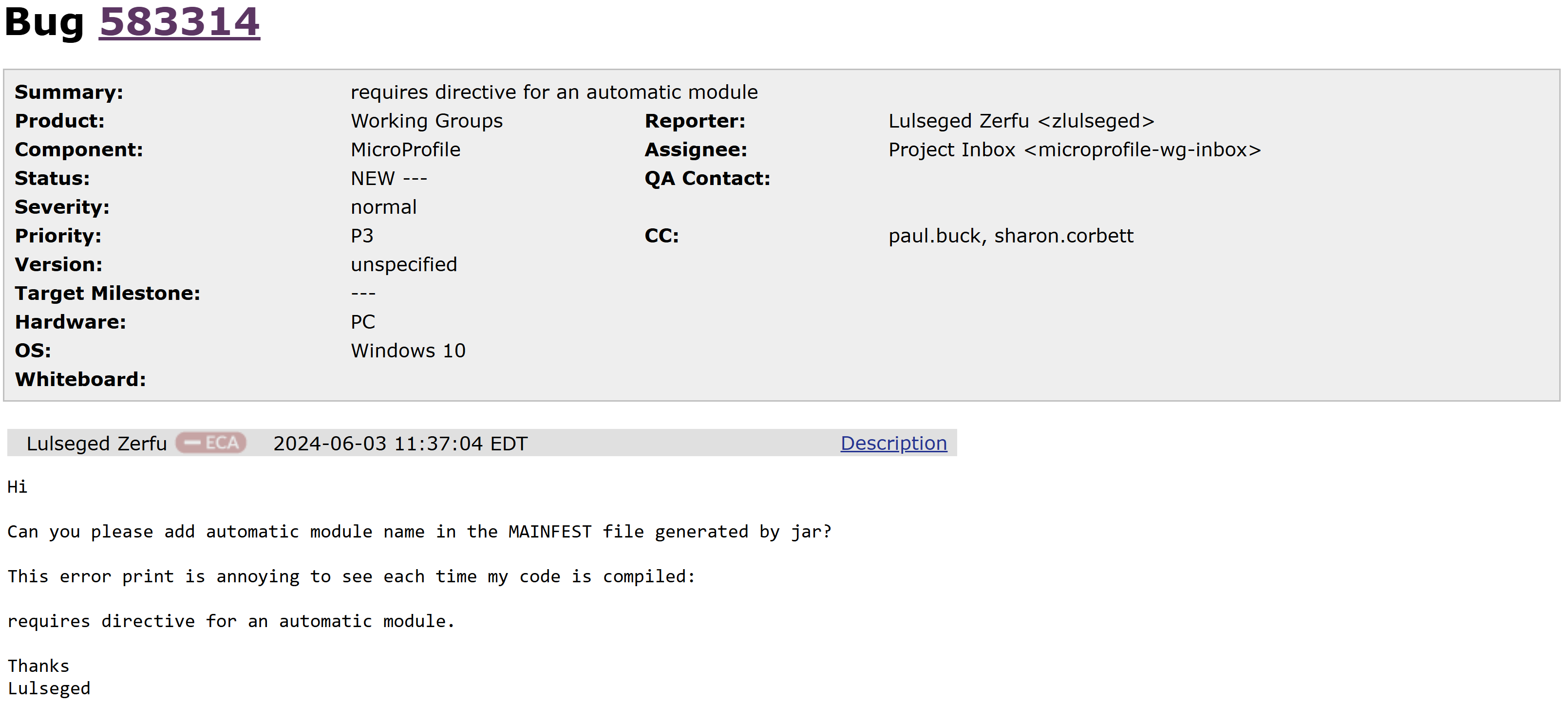}
    \caption{Sample Bug Report}
    \label{fig:enter-label}
\end{figure}

\subsection{Natural Language Processing}
Software development projects often include open bug repositories. Natural Language Processing (NLP) is a part of computer science that works to help computers better understand human written or spoken language\cite{NLP+1994}. We used natural language processing techniques to analyze the bug reports. Before we input the bug reports into a neural network, we needed to preprocess the reports. As shown in Figure 2, before we can analyze the text, we have to process it. This included tokenization, fixing spelling, removing stop words, and stemming. We used the package NLTK to do this \cite{loper+2002}. In our tokenization step, we separated the words in the reports to make each word a token and lowercase. We removed stop words, which are words that are used often but do not add value to the sentence,s such as "the", "I", or "on". This allowed us to analyze only the significant words that were added to the report. Stemming of the tokens involves changing words to the core part to include more words from a sentiment analysis dictionary. An example of stemming is changing "advancement" to "advance" \cite {singh+2017}. After preprocessing bug reports, we used sentiment analysis and the SentiWordNet lexicon \cite{baccianella+2010}. This dictionary assigned the words it had matches for a positive, negative, and objective (neutral) score. 

\subsection{BERTopic}
BERTopic \cite{grootendorst+2022} is an advanced topic modeling technique that uses Bidirectional Encoder Representations from Transformers (BERT) to generate topics from textual data and bug report descriptions. First, we had to pre-train it on a training set so that it could create embeddings that attempt to capture the meaning of the text. It then groups bugs that have similar embeddings and generates topics. 

\section{Related Work} \label{related-work}
Automated bug triage in open-source software systems has been studied over the past decades (e.g., see \cite{MoinKhansari2010,MoinNeumann2012}).

\subsection{Sentiment Analysis in Bug Reports}
Ramay et al. \cite{ramay+2019} used sentiment analysis to predict whether a bug was severe or not severe by using the Senti4SD lexicon and a Convolutional Neural Network (CNN). Found that neural networks with sentiment analysis outperformed the leading machine learning algorithms Multinomial Naïve Bayes (MNB) and Random Forest (RF). With F-Score of 0.8436, 0.7818, and 0.8007.
   
Baarah et al. 2021 \cite{baarah+2021} used sentiment analysis and found that Multilayer Perceptron (MLP) predicts severity better than the lexicon-based neural network. F-Score of 0.81 compared to 0.52. For their sentiment analysis, they took the difference between the positive score and the negative score of a report. If the difference was positive, it was labeled as a positive report. If the difference is negative, they labeled it a negative report before testing the neural network's success in predicting severity.
    
Umer et al. \cite{Umer+2020} used a Convolutional Neural Network (CNN) to prioritize bugs after calculating the sum of positive and negative emotion scores to get a total emotionality score. They inputted that sum and the feature vectors of each bug report into a CNN and got an F-Score of 0.6364.

We used these three sources as ideas on how to conduct sentiment analysis on bug reports because they had been proven to be successful in predicting other aspects of bug reports, such as priority and severity. We took inspiration to use a CNN model and an MLP model \cite{Umer+2020} \cite{baarah+2021}.

\subsection{Time-to-Fix}
Because there are so many ways to categorize time to fix as either numerical or categorical, it was very hard to compare metrics across different prediction techniques. Another challenge was that different platforms for bug reports included unique information about each report. In this section, we will first discuss the papers that have the highest F1 scores when predicting the time to fix, and then we will discuss the papers that are working with similar datasets and use similar metrics for short and long. 

Habayeb et al. \cite{habayeb+2018} used a Hidden Markov Model (HMM) and found good results when looking at a Firefox project and used the temporal dataset of activities that occurred during the life cycle of a bug.

Ardimimento in 2024 \cite{ardimento+2024} had successful results using Long Short Term Memory (LSTM) to predict the time to fix. They only looked at bugs that had been fixed. The dataset they used was an Eclipse project, and both examined the chronological order in which bug reports are fixed using temporal sequences. However, our dataset does not include information on the activities or steps taken to fix a bug. So our project is more inclusive to those bug reports that have not been completed yet and can better be applied to bug reports as they come in. 

Ardimento in 2020 \cite{ardimento+2020} was successful when using Bert to predict bug fixing time. This model also only looked at bug reports that were eventually labeled fixed. They labeled the longest 50\% of bugs as long and the rest as short. 

Marks et al. \cite{marks+2011} looked at an Eclipse project similar to ours and used information about the bug report that can be found before it gets fixed such as reporter, location (component), and description. They used a random forest classifier and got an accuracy of 0.65.

Ardimento in 2017 \cite{ardimento+2017} used knowledge extraction and looked at data sets from LiveCode, Novell, and OpenOffice to predict the time to fix. They split up the data 75\% of bugs labeled as fast and 25\% labeled as slow. This was similar to how we split our data into slow and fast categories. They used features that can be created or found before the bug is fixed such as description, comments from the developer, priority, severity, and more. They got the best results using the LiveCode dataset with an accuracy of 0.52, a recall rate of 0.62, and a precision of 0.22. Regardless that this is a different dataset we will compare our results with these results because they use features provided early in the triaging process.

Panjer \cite{panjer+2007} examined the time to resolution of all bugs instead of the time to fix. They used data mining techniques to get an accuracy of 0.34. 

These two sources had similar information in their data sets as we had in ours, so we found comparing our work to theirs to be the closest comparison.

\subsection{Time-to-Resolution}
One of the things this paper focused on is correctly using the term time-to-fix for bug reports and how that is compared to time-to-resolution. Time-to-fix is strictly for bugs that have been fixed and predicting how long those will take to fix. Time-to-resolution is more useful and should be labeled as such. Papers often \cite{ardimento+2020} \cite {ardimento+2017} use them interchangeably and talk about how they use Bert to predict bug resolution time, but in their proposed approach, they sort out only the bugs that say fixed as a resolution. We suggest a more holistic approach when referring to either or and being consistent across papers. 

\section{Proposed Approach} \label{proposed-approach}
We propose a practical and comprehensive application of sentiment analysis to predict various classifications within a bug report dataset. Specifically, we examine the emotional content of each bug report to forecast its eventual resolution outcome. Our approach employs three predictive models. First, we estimate the time-to-resolution, which categorizes whether developers will close a bug with a resolution label, regardless of whether it was actually fixed. Second, we predict the time-to-fix, distinguishing whether the bug will ultimately be resolved. Finally, we attempt to determine the specific resolution label assigned to the bug, which we refer to as its "destiny."

Each predictive model follows a consistent framework with slight modifications to the model architecture and input features. We utilize the Eclipse Bugzilla dataset \cite{bughub+2018}, which contains 85,156 bug reports, each devoid of developer comments and relying solely on reporter descriptions. Since the dataset did not originally include time-to-resolution information, we computed duration labels by subtracting the reported date from the resolution date. To facilitate classification, we created a binary time label: bugs with resolution times in the lowest 70\% were labeled "short," while those in the top 30\% were labeled "long." Figure 2 illustrates the distribution of bug resolution times.

Our preprocessing pipeline incorporates additional labels to quantify both the duration of bug activity and the emotional tone of the bug description. We employ sentiment analysis using the SentiWordNet lexicon \cite{baccianella+2010}. To the best of our knowledge, previous research has not leveraged sentiment analysis to predict time-to-resolution or time-to-fix. Additionally, we remove stop words and apply lemmatization to enhance the sentiment analysis process. A separate "destiny" column is introduced to indicate whether the bug will ultimately be fixed, distinct from the resolution label prediction. Other data refinements include converting priority levels into numerical values to facilitate analysis.

The sentiment analysis specifically introduces multiple new columns to the dataset, many of which are used in order to make future classifications. The process itself creates two new columns 'pos\_score' and 'neg\_score,' which store numeric values that indicate how positive and how negative the test input is. Any given text can have both positive and negative elements at the same time. Using these two columns we create another column known as 'Emotion,' which is used to assist all of our models in their classifications. This specific column indicates if a specific text is overall positive or overall negative by subtracting the 'Neg\_Score' from the 'Pos\_Score'. Next, we also use the same positive and negative columns to create an 'Emotionality' column, which is also used in all of our models for their classification. This specific column is the combined total of the positive and negative emotions in the given text essentially showing the magnitude of emotion.

When splitting the dataset we employ an 80\% training and 20\% testing split. This specific split is not done randomly and instead places the oldest bugs in the training in an attempt to predict based on bugs not yet seen to prove the effectiveness of the model in a real-world application. This specific split style does result in a disparity between the 'resolution' label in the dataset, which introduces a new resolution label in the testing set that was not present in the training set. We choose to ignore this new label as the current split is not viable for predicting any outcome based on the specific label.

Beyond preprocessing, we standardize our machine learning models. For each prediction task, we apply four machine learning models, including multi-layer perceptron (MLP) networks and one-dimensional convolutional neural networks (CNNs). Additionally, we implement a BERTopic model to extract topics from bug descriptions. This model, fine-tuned on a training subset of the Eclipse Bugzilla dataset, identifies 20 distinct topics. We then use this trained model to assign topics to each bug report \cite{grootendorst+2022}. Each prediction model considers features such as emotion, emotionality, and priority, with BERT serving as a comparative baseline. Furthermore, we introduce model variants that balance input weights to improve classification outcomes.

The first major predictive task involves time-to-resolution, which determines when a bug receives a resolution label, independent of whether the issue was genuinely fixed. This task is framed as a binary classification problem, distinguishing between "short" and "long" resolution times. Additionally, we also treat this as a regression problem and make specific time guesses for the predicted number of hours until the bug is resolved. For this we employ CNN and logistic regression models and do not use BERT. We also attempt to use these models to bias based on our 'long' and 'short' labels as well the entire dataset to compare the outcomes and the reliability of our prediction for the specific tasks.

The second task, time-to-fix prediction, builds upon the time-to-resolution model but specifically predicts when a bug will be fixed rather than merely resolved. Like the previous model, this task also employs binary classification and a regression problem where we attempt to predict the exact time frame to fix the bug.

The final predictive task involves determining the eventual resolution label assigned to a bug. This is formulated as a multi-class classification problem, predicting among seven possible resolution labels: 'WONTFIX' 'DUPLICATE' 'FIXED' 'WORKSFORME' 'NDUPLICATE' 'INVALID'
 'NOT\_ECLIPSE'. The same input features and machine learning models used in the previous tasks are applied to this classification problem, but the model itself makes a multi-class classification.


\begin{figure}[h]
    \centering
    \includegraphics[width=0.8\linewidth]{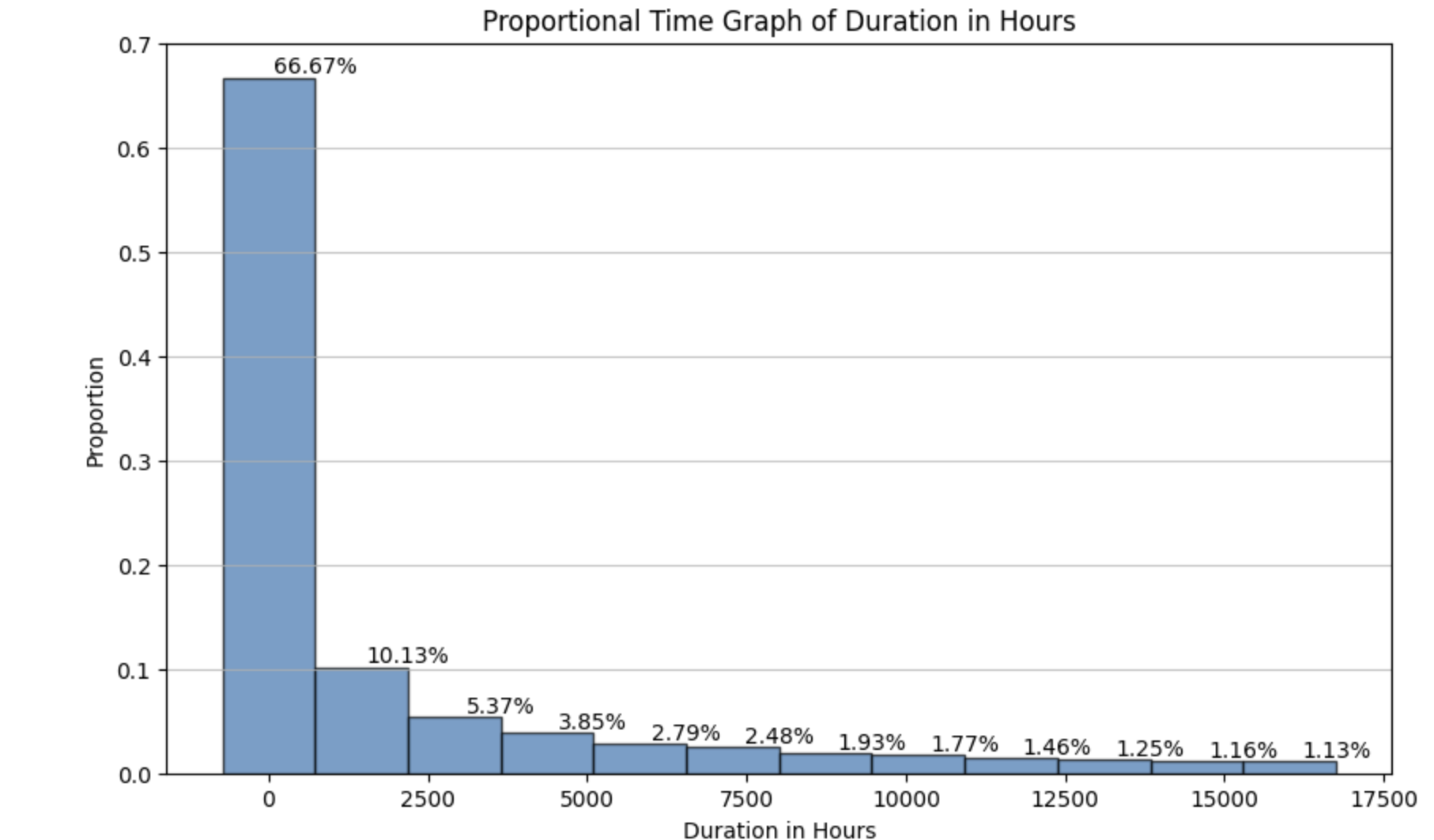}
    \caption{Distribution of Time to Resolution}
    \label{fig:enter-label}
\end{figure}
\begin{figure}[h]
    \centering
    \includegraphics[scale=0.3]{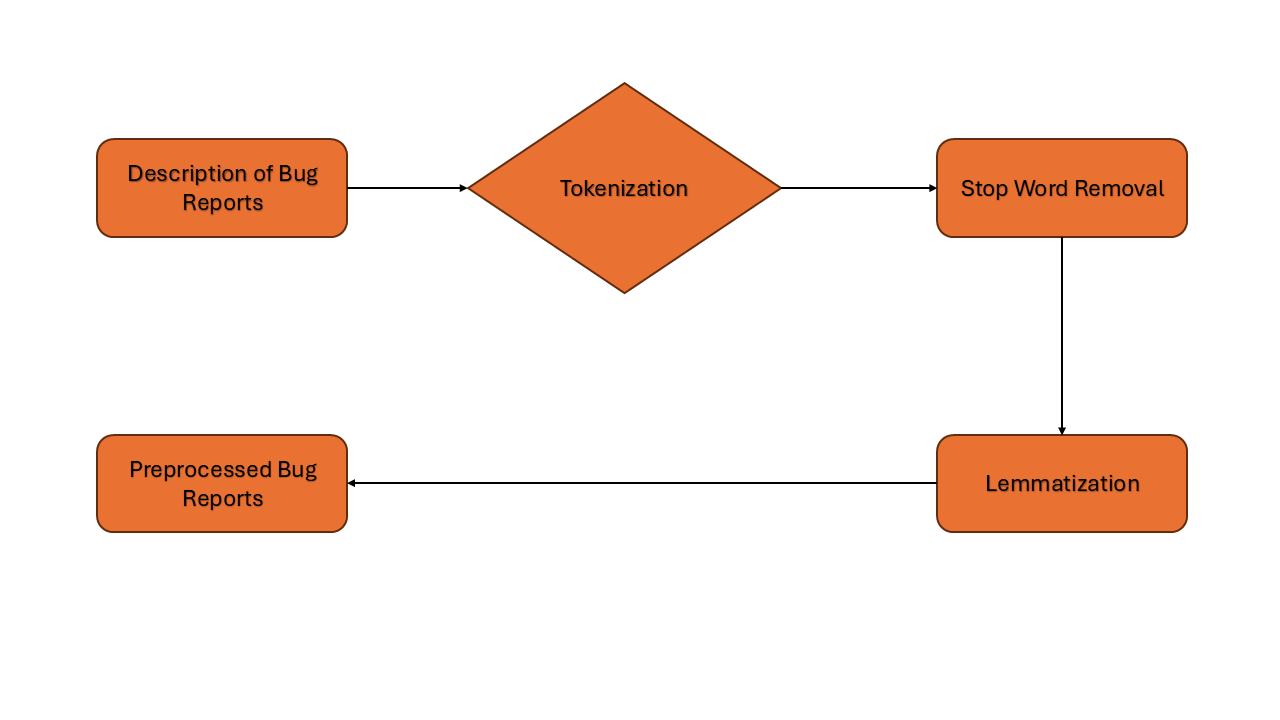} 
    \caption{Preprocessing of bug reports}
    \label{fig:enter-label}
\end{figure}


\begin{figure}[h]
    \centering
    \includegraphics[width=0.5\linewidth]{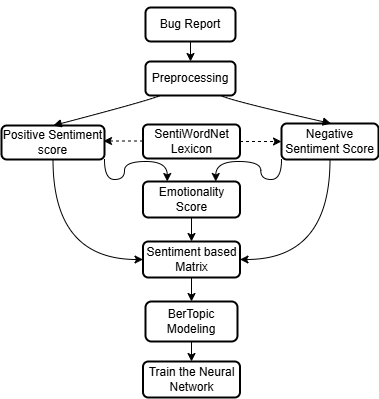}
    \caption{High Level Pipeline}
    \label{fig:enter-label}
\end{figure}

Unlike the time-to-fix prediction papers, we did not sort out the bug reports that had been labeled fixed in the resolution column. After we had completed all of these different features, we then used them through two different neural networks: a Convolution Neural Network (CNN) and a multi-layer perceptron Model (MLP). We compared how different models did with different features such as unique combinations of Emotion, Emotionality, Priority, and predicted topic. We measured these different models by using weighted recall scores, precision scores, F1 scores, and accuracy. We used weighted scores for each metric to account for differences in class distribution. A high recall score means that the model correctly identifies many positive instances. 
\[
\text{Recall}_i = \frac{\text{True Positives (TP)}_i}{\text{True Positives (TP)}_i + \text{False Negatives (FN)}_i}
\]
A high recall measure implies that the model correctly identifies most of the positive instances.

\[
\text{Precision}_i = \frac{\text{True Positives (TP)}_i}{\text{True Positives (TP)}_i + \text{False Positives (FP)}_i}
\]
A high-precision measure implies that if the model guesses a positive instance, it is very likely to be one.

\[
\text{F1-Score}_i = 2 \cdot \frac{\text{Precision}_i \cdot \text{Recall}_i}{\text{Precision}_i + \text{Recall}_i}
\]
The F1 score is a combined metric of precision and recall.

\[
\text{Weighted Recall} = \sum_{i=1}^{N} \frac{T_i}{T} \cdot \text{Recall}_i
\]
\[
\text{Weighted Precision} = \sum_{i=1}^{N} \frac{T_i}{T} \cdot \text{Precision}_i
\]
\[
\text{Weighted F1-Score} = \sum_{i=1}^{N} \frac{T_i}{T} \cdot \text{F1-Score}_i
\]
Where \( T_i \) is the number of true instances in class \( i \), and \( T \) is the total number of instances across all classes.

Accuracy is the amount of bug reports that the model labeled correctly as short or long.

\section{Experimental Results} \label{experimental-results}
We tried various approaches with the models used in each classification problem. When displaying the results in our tables we will display the weighted average for every value comparison as well as the accuracy. Our evaluation metrics will be scored as a decimal value with 1 being the highest or equivalent to 100\% when considering accuracy.
\subsection{Predicting Time Label}
\subsubsection{Time-to-Resolution}
When making our binary prediction between short and long for the time-to-resolution we observe equal performance between the CNN and MLP models for the unbalanced models. When both using BERT and the normal parameters the model saw an accuracy of 79\%. Next we also balanced the dataset before sending it as an input to the model in order to balance the weight of the short and long classes. This classification resulted in reduced overall accuracy, with the best accuracy with this approach being 68\%. In this case BERT improved the results of the model and resulted in a much better accuracy. This specific weighted approach saw a better classification for the 'long' classification than its non-weighted counterpart.
\begin{table}[h!]
\centering
\caption{Comparison between time-to-resolution models}
\begin{tabular}{|>{\centering\arraybackslash}p{3cm}|c|c|c|c|}
  \hline
  Model & Precision & Recall & F1 Score & Accuracy \\
  \hline
  MLP (Emotion, Emotionality, Priority) & 0.75 & 0.79 & 0.70 & 0.79 \\
  \hline
  CNN (Emotion, Emotionality, Priority) & 0.75 & 0.79 & 0.70 & 0.79 \\
  \hline
  MLP (Emotion, Emotionality, Priority, Predicted Topic) & 0.75 & 0.79 & 0.70 & 0.79 \\
  \hline
  CNN (Emotion, Emotionality, Priority, Predicted Topic) & 0.75 & 0.79 & 0.70 & 0.79 \\
  \hline
  MLP (Emotion, Emotionality, Priority) Weighted & 0.67 & 0.59 & 0.62 & 0.59 \\
  \hline
  CNN (Emotion, Emotionality, Priority) Weighted & 0.67 & 0.58 & 0.61 & 0.58 \\
  \hline
  MLP (Emotion, Emotionality, Priority, Predicted Topic) Weighted & 0.67 & 0.60 & 0.63 & 0.60 \\
  \hline
  CNN (Emotion, Emotionality, Priority, Predicted Topic) Weighted & 0.67 & 0.68 & 0.67 & 0.68 \\
  \hline
\end{tabular}
\label{table:1}
\end{table}
\subsubsection{Time-to-fix}
We used the same process for predicting time-to-fix that was used for time-to-resolution. The only difference being that we applied an additional step of preprocessing to the dataset that restricted the dataset to only contain bugs that are marked as 'FIXED'. The result of this process is a classification accuracy of 79\%. With the balanced inputs the best accuracy obtained in this case was ~64\% with a basic MLP model without the BERT topic modeling.
\begin{table}[h!]
\centering
\caption{Comparison between time-to-fix models}
\begin{tabular}{|>{\centering\arraybackslash}p{3cm}|c|c|c|c|}
  \hline
  Model & Precision & Recall & F1 Score & Accuracy \\
  \hline
  MLP (Emotion, Emotionality, Priority) & 0.78 & 0.79 & 0.71 & 0.79 \\
  \hline
  CNN (Emotion, Emotionality, Priority) & 0.78 & 0.79 & 0.71 & 0.79 \\
  \hline
  MLP (Emotion, Emotionality, Priority, Predicted Topic) & 0.78 & 0.79 & 0.71 & 0.79 \\
  \hline
  CNN (Emotion, Emotionality, Priority, Predicted Topic) & 0.78 & 0.79 & 0.71 & 0.79 \\
  \hline
  MLP (Emotion, Emotionality, Priority) Weighted & 0.68 & 0.64 & 0.66 & 0.64 \\
  \hline
  CNN (Emotion, Emotionality, Priority) Weighted & 0.68 & 0.58 & 0.62 & 0.58 \\
  \hline
  MLP (Emotion, Emotionality, Priority, Predicted Topic) Weighted & 0.69 & 0.58 & 0.62 & 0.58 \\
  \hline
  CNN (Emotion, Emotionality, Priority, Predicted Topic) Weighted & 0.68 & 0.53 & 0.58 & 0.53 \\
  \hline
\end{tabular}
\label{table:1}
\end{table}
\subsection{Predicting Numeric Time}
The numeric time classification was done as a regression problem. We tried different approaches for this specific classification. The initial tests did not perform well in this regard, which meant that we needed to diversify our approach. Our best results on the entire dataset for time-to-resolution was a mean absolute error of 2890 hours. Then, after restricting the dataset to only classify the 'short' labeled data, the best result observed was a mean absolute error of 575 hours. Our results for time-to-fix differed slightly with a mean absolute error of 2422 for the full dataset and 662 for the 'short' label only. Our results for the 'long' labeled data were overall worse, with the time-to-resolution having a mean absolute error of 7769 and time-to-fix having 5170, respectively.
\begin{table}[h!]
\centering
\caption{Comparison between numeric time models for Time-to-Resolution}
\begin{tabular}{|>{\centering\arraybackslash}p{3cm}|c|c|c|c|}
  \hline
  Model & Mean Absolute Error & Mean Squared Error \\
  \hline
  CNN (Full Dataset) & 2,890 & 42,422,784 \\
  \hline
  CNN (Short) & 575 & 983,709 \\
  \hline
  CNN (Long) & 7,769 & 116,059,196 \\
  \hline
  Linear Regression (Full Dataset) & 2,893 & 42,445,410 \\
  \hline
  Linear Regression (Short) & 578 & 1,029,487 \\
  \hline
  Linear Regression (Long) & 8,835 & 115,698,132 \\
  
  \hline
\end{tabular}
\label{table:1}
\end{table}
\begin{table}[h!]
\centering
\caption{Comparison between numeric time models for Time-to-Fix}
\begin{tabular}{|>{\centering\arraybackslash}p{3cm}|c|c|c|c|}
  \hline
  Model & Mean Absolute Error & Mean Squared Error \\
  \hline
  CNN (Full Dataset) & 2,422 & 28,267,867 \\
  \hline
  CNN (Short) & 662 & 1,044,963 \\
  \hline
  CNN (Long) & 5,170 & 63,181,711 \\
  \hline
  Linear Regression (Full Dataset) & 2,427 & 28,541,468 \\
  \hline
  Linear Regression (Short) & 671 & 1,144,002 \\
  \hline
  Linear Regression (Long) & 6,125 & 60,252,990 \\
  
  \hline
\end{tabular}
\label{table:1}
\end{table}
\subsection{Predicting Resolution Label (Destiny)}
The resolution label prediction is a multi-class classification. One of the eventual labels is removed from the dataset due to the lack of the labels' presence in both the test and train set. The model in this case achieved an accuracy of 57\% but fail to truly classify any labels other than fixed with a reasonable accuracy.
\begin{table}[h!]
\centering
\caption{Comparison between Destiny models}
\begin{tabular}{|>{\centering\arraybackslash}p{3cm}|c|c|c|c|}
  \hline
  Model & Precision & Recall & F1 Score & Accuracy \\
  \hline
  MLP (Emotion, Emotionality, Priority) & 0.34 & 0.57 & 0.42 & 0.57 \\
  \hline
  CNN (Emotion, Emotionality, Priority) & 0.33 & 0.57 & 0.42 & 0.57 \\
  \hline
  MLP (Emotion, Emotionality, Priority, Predicted Topic) & 0.34 & 0.57 & 0.42 & 0.57 \\
  \hline
  CNN (Emotion, Emotionality, Priority, Predicted Topic) & 0.33 & 0.57 & 0.42 & 0.57 \\
  \hline
  MLP (Emotion, Emotionality, Priority) Weighted & 0.46 & 0.15 & 0.18 & 0.15 \\
  \hline
  CNN (Emotion, Emotionality, Priority) Weighted & 0.33 & 0.57 & 0.42 & 0.57 \\
  \hline
  MLP (Emotion, Emotionality, Priority, Predicted Topic) Weighted & 0.48 & 0.13 & 0.15 & 0.13 \\
  \hline
  CNN (Emotion, Emotionality, Priority, Predicted Topic) Weighted & 0.33 & 0.57 & 0.42 & 0.57 \\
  \hline
\end{tabular}
\label{table:1}
\end{table}

We also ran experiments using a Support Vector Machine as a regression model to find if there was a relationship between emotionality and time-to-resolution. We created a scatter plot of this data to visualize it before running the machine learning model, as seen in Figure 5. We found a \( R^2 \)=-0.23.
\begin{figure}[h]
    \centering
    \includegraphics[width=0.9\linewidth]{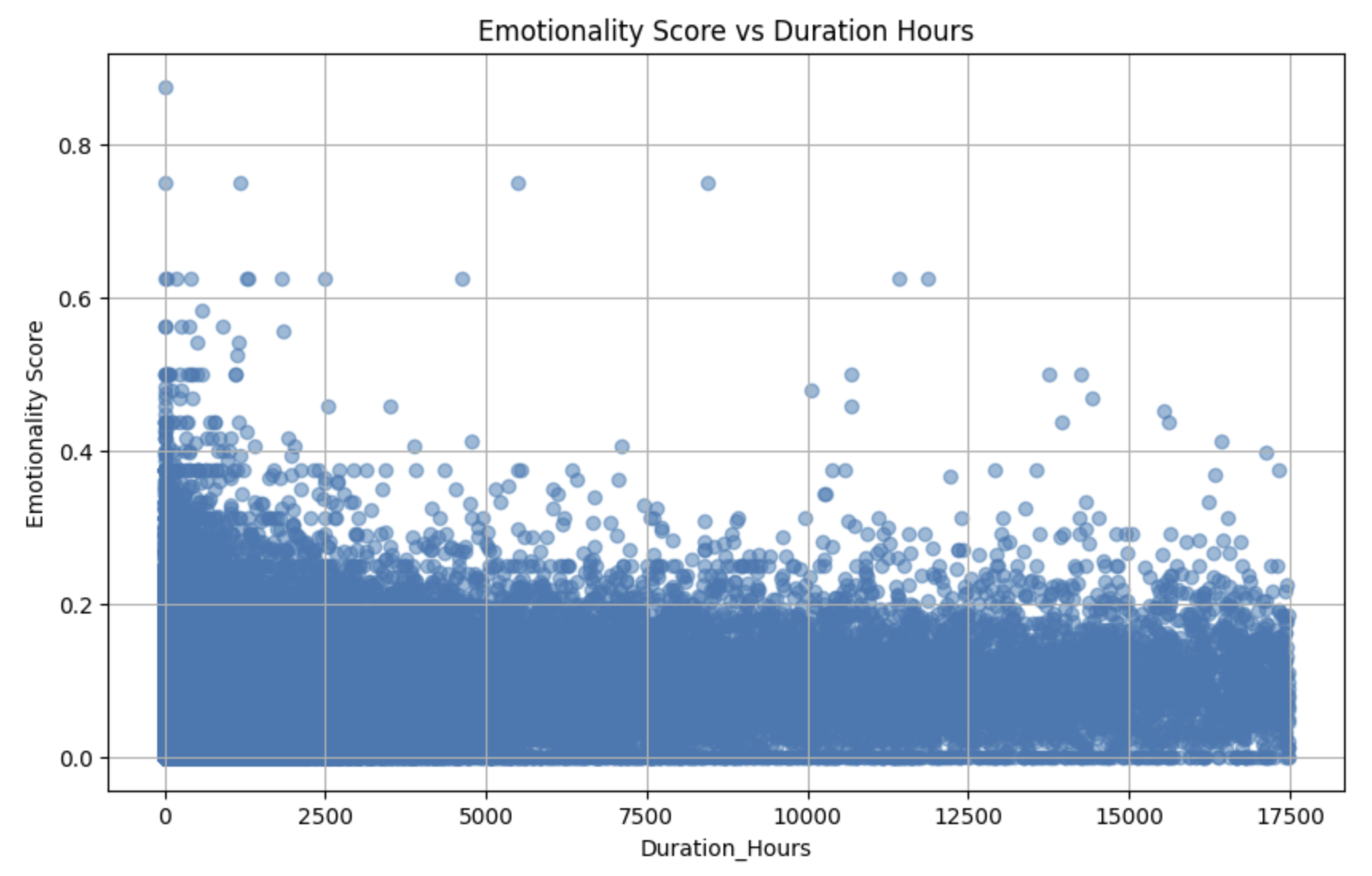}
    \caption{Scatter plot of Emotionality Score's Relationship with Time to Resolution}
    \label{fig:enter-label}
\end{figure}

\section{Discussion} \label{discussion}
All of the models performed with a high overall accuracy when performing the binary and multi-class classifications.  However in spite of the overall decent accuracy across all models the performance of each variant of our approach, i.e., cnn vs mlp is either the same or extremely similar. As for BERT it was used with inconsistent results; in some cases, such as with the weighted CNN approach on time-to-resolution, BERT was able to greatly improve our results and make the otherwise less reliable weighted version more practical for general use. The non-weighted models, while having a higher accuracy, show that they rarely predict the 'long' classification. Thus we introduced more balancing to the model's inputs, which greatly improved the 'long' classification results while reducing the overall performance of most of the other models. The difference between the performance of the models between time-to-fix and time-to-resolution were minor, and the model proves that it is viable in both circumstances. With time-to-fix, the non-weighted models had the same accuracy, but time-to-fix had slightly better overall results, with its F1 score being slightly higher than the time-to-resolution. 

With the exact numeric time predictions, both of the models used performed similarly with each being slightly better than the other in one of our two metrics. The best overall performance that was achieved was a mean absolute error of 575 for the prediction of the 'short' only label. Seemingly making an exact prediction for the time that a bug will take to be resolved is impractical when applied to a dataset with as much variance as ours. Instead, predicting our "short" time frame produced more reasonable and reliable results while still encompassing 70\% of the total dataset. Our 'long' classification, in this case, performed worse overall than using the full dataset, but this is likely due to the 'short' accuracy bringing down the average. The difference between the time-to-resolution and the time-to-fix results is overall not very large, with the time-to-fix seemingly performing better for the 'long' and full dataset predictions but performing worse for the 'short' predictions. Each time these models are run, the results change slightly, meaning that performance is instead close to being standard across all of the problems and models. Essentially, all models run have a slight variance in their output, with the results of all models being numerically close to one another. 

The destiny prediction itself saw an overall low accuracy with the best performing models only achieving 57\% accuracy. These models only predicted the 'FIXED' and 'WONTFIX' data labels and were not able to distinguish any of the other possible classifications. By adding balancing to the input of the dataset using SMOTE, we managed to have the model predict at least partly the other possible classification labels. This however saw an extreme decrease in overall accuracy with the best models struggling to exceed 15\% accuracy. Additionally, for the weighted CNN destiny predictions, the weighting did not have any effect on the dataset. This potentially hints at possible errors in the code, but we were unable to resolve these. This results in the weighted and non-weighted models having the exact same accuracy and results.

It is believed that the weighted approaches, while not as accurate, are more valuable for the classification problem rather than the non-weighted input approaches. We noticed that while the non-weighted approaches boast a high accuracy most of this is weighted toward the short label, which exists as 70\% of our dataset. In a practical setting it would instead be more valuable to know if a bug will take an exponentially long amount of time to be resolved; thus, having a slightly more reliable long classification is more beneficial. We also observed that for the destiny prediction, our model was primarily predicting the 'fixed' classification with a high rate of accuracy, but it was unable to predict the other classification labels without balancing. For this specific problem, knowing if a bug will end up being fixed is highly valuable, more so than the potential other labels, but having access to a variant of the model that specifically targets the other labels could be significant depending on the use case.

The use of BERT was met with unreliable results. There are some cases where it made no change to the output, others where is made improvements, and some where it made the results worse. Overall, it could be useful when applied to this specific classification problem, but ultimately, with the simple BERT model employed in this research the results are rather minimal and unreliable.
\begin{table}[h!]
\centering
\caption{Comparison of Model Performance Metrics}
\begin{tabular}{|>{\centering\arraybackslash}p{3cm}|c|c|c|c|}
  \hline
  Model & F1 Score & Recall & Precision & Accuracy \\
  \hline
  HMM \cite{ardimento+2024} & 0.70 & 0.69 & 0.69 & 0.71 \\
  \hline
  DEEPLSTM \cite{sepahvand+2020} & 0.96 & 0.87 & 0.79 & 0.88 \\
  \hline
  LSTM \cite{ardimento+2024} & 0.97 & 0.88 & 0.79 & 0.87 \\
  \hline
  CNN (Emotion, Emotionality, Priority, Predicted Topic) & 0.71 & 0.79 & 0.78 & 0.79 \\
  \hline
\end{tabular}
\label{table:1}
\end{table}
To the best of our knowledge, Table Seven shows how our work compares to the state-of-the-art approaches for predicting bug-fixing time using information from before the bug was fixed.
\begin{table}[h!]
\centering
\caption{Comparison of Model Performance Metrics Using Features From Before a Bug is Resolved}
\begin{tabular}{|>{\centering\arraybackslash}p{3cm}|c|c|c|c|}
  \hline
  Model & F1 Score & Recall & Precision & Accuracy \\
  \hline
  Data Mining \cite{panjer+2007} & -- & -- & -- & 0.34 \\
  \hline
  Random Forest\cite{marks+2011} & -- & -- & -- & 0.65 \\
  \hline
  Knowledge Extraction \cite{ardimento+2017} & -- & 0.62 & .22 & 0.52 \\
  \hline
  CNN (Emotion, Emotionality, Priority, Predicted Topic) & 0.71 & 0.79 & 0.78 & 0.79 \\
  \hline
\end{tabular}
\label{table:1}
\end{table}
All of these approaches except Panjer \cite{panjer+2007} look specifically at time-to-fix instead of resolution. Because our model works even better for time-to-fix, we achieved state-of-the-art results using BERTopic and sentiment analysis.

\subsection{Threats to Validity} \label{threats-to-validity}
It is hard to compare our results to other approaches because we used a different dataset that did not include developer comments and may be formatted differently and include more or less information than our dataset. Also, the metrics for labeling bug reports as long or short are not consistent with other work done on bug-fixing time. Our code also shows an irregularity in our weighted CNN model, showcasing no change after balancing is done.

\section{Conclusion and Future Work} \label{conclusion-future-work}
We have found a relationship between how emotional a description report is and how long it takes to resolve or fix. Further work could include exploring if using emotionality as an added feature in other models can improve them, as we did with BERTopic, and continuing to explore how sentiment analysis can help with time prediction.  Another example of possible future work is looking at whether emotionality's effect on time to resolution is dependent on whether the bug was reported by a user or a developer. Overall we were successful in predicting bug reports' time-to-resolution using sentiment analysis and BERTopic.

\section*{Software and Data Availability}
The prototype is available under a permissive open-source license at \url{https://github.com/qas-lab/PopeREU}. The data used for the evaluation are also publicly available at \url{https://github.com/logpai/bughub/tree/master/EclipsePlatform}.

\section*{Acknowledgment}
This material is based upon work supported by the U.S. National Science Foundation (NSF) under Grant No. 2349452. Any opinions, findings, conclusions, or recommendations expressed in this material are those of the authors and do not necessarily reflect the views of the NSF.

\bibliography{refs}
\bibliographystyle{IEEEtran}

\end{document}